\documentclass[a4paper]{jpconf}
\usepackage{graphicx}

\newcommand{\Tss}[3]{ \ensuremath{#1 _{\mathrm{#2}} ^{\mathrm{#3}} }}
\newcommand{\Tsub}[2]{\ensuremath{#1_{\mathrm{#2}}}}

\newcommand{\Tk}[1]{\Tsub{\kappa}{#1}}

\newcommand{\Tkax}[1]{\ensuremath{\kappa _{#1}}}
\newcommand{\Tl}[1]{\Tsub{l}{#1}}
\newcommand{\Tv}[1]{\Tsub{v}{#1}}

\newcommand{\Tt}[1]{\Tsub{\tau}{#1}}

\newcommand{\TDebye}{\Tsub{\mathit{\Theta}}{D}}

\newcommand{\TkB}{\Tsub{k}{B}}
\newcommand{\TJ}[2]{\Tss{J}{#1}{#2}}

\newcommand{\TSCO}[3]{Sr$_{#1}$Cu$_{#2}$O$_{#3}$}

\newcommand{\Tion}[2]{\ensuremath{\mathrm{#1}^{\mathrm{#2}}}}
\newcommand{\TCuII}{Cu$^{2+}$}
\newcommand{\TPdII}{\Tion{Pd}{2+}}

\newcommand{\Tm}{\ensuremath{ \mathrm{m} }}

\newcommand{\Tsec}{\ensuremath{ \mathrm{s} }}

\newcommand{\Tdeg}{\ensuremath{\mathrm{\char'27 \kern-.2em}}}
\newcommand{\TdegC}{\ensuremath{\mathrm{\char'27 \kern-.2em \hbox{C}}}}
\newcommand{\TK}{\ensuremath{ \mathrm{K} }}
\newcommand{\diff}{\ensuremath{\mathrm{d}}}

\begin{document}
%
%
\title{Enhancement of Thermal Conductivity due to Spinons in the One-Dimensional Spin System SrCuO$_2$}
\author{
  T~Kawamata$^{1}$, 
  N~Kaneko$^{2}$, 
  M~Uesaka$^{2}$, 
  M~Sato$^{2}$, 
  Y~Koike$^{2}$
}
\address{$^1$ Nishina Center for Accelerator-Based Science, RIKEN, 2-1 Hirosawa, Wako 351-0198, Japan}
\address{$^2$ Department of Applied Physics, Tohoku University, 6-6-05 Aoba, Aramaki, Aoba-ku, Sendai 980-8579, Japan}
\ead{tkawamata@riken.jp}
%
%
\begin{abstract}
We have measured the thermal conductivity along the $c$-axis parallel to the spin-chains, \Tkax{c}, of the one-dimensional antiferromagnetic spin system SrCuO$_2$, using as-grown and O$_2$-annealed single-crystals grown from raw materials with 99.9~\% (3N) and 99.99~\% (4N) purity. 
The value of \Tkax{c} around 50~K, where large contribution of the thermal conductivity due to spinons, \Tk{spinon}, is observed, is markedly enhanced by both the increase of the purity of raw materials and the O$_2$-annealing. 
Therefore, the increase of \Tkax{c} implies that \Tk{spinon} is enhanced due to the decrease of spin defects caused by impurities in raw materials and by oxygen defects. 
The mean free path of spinons is as large as about 24000~{\AA} at low temperatures in the O$_2$-annealed single-crystal grown from raw materials with 4N purity. 
\end{abstract}
%
%
\section{Introduction}
Recently, high thermal conductivity due to magnetic excitations in low-dimensional quantum spin systems has attracted great interest.  
For one-dimensional (1D) antiferromagnetic (AF) integrable Heisenberg systems with the spin quantum number $S=1/2$, some theoretical calculations have suggested that the thermal conduction due to magnetic excitations, called spinons, tends to be ballistic, leading to large values of thermal conductivity due to spinons, \Tk{spinon} \cite{Castella:PRL74:1995:972,Zotos:PRB55:1997:11029,Klumper:JPA35:2002:2173}. 
Actually, large values of \Tk{spinon} have been observed in the $S=1/2$ 1D AF system \TSCO{2}{}{3} \cite{Sologubenko:PRB62:2000:R6108,Sologubenko:PRB64:2001:054412} and \TSCO{}{}{2} \cite{Sologubenko:PRB64:2001:054412}. 
Our recent work has proved that the thermal conduction due to spinons in \TSCO{2}{}{3} is ballistic at low temperatures \cite{Kawamata:JPSJ77:2008:034607}. 
In this study, the mean free path of spinons, \Tl{spinon}, 
of Sr$_2$Cu$_{1-x}$Pd$_x$O$_3$, 
where {\TCuII} ions with $S=1/2$ are partially replaced by nonmagnetic {\TPdII} ions, 
has been found to be comparable with the length between spin defects estimated from the magnetic susceptibility measurements at low temperatures. 
This means that the spinons carry heat between spin defects without being scattered. 
In other words, there is a possibility of the enhancement of \Tk{spinon} if the number of spin defects decreases in 1D AF systems. 

So far, the highest value of \Tk{spinon} among 1D AF systems has been obtained in \TSCO{}{}{2} \cite{Sologubenko:PRB64:2001:054412} 
with zigzag spin-chains formed from two CuO chains combined with each other by sharing their edges of CuO$_4$ squares. 
In \TSCO{}{}{2}, the exchange interaction between the nearest Cu spins with the 180{\Tdeg} Cu-O-Cu bond, \TJ{}{}, 
and that between the diagonal Cu spins with the nearly 90{\Tdeg} Cu-O-Cu bond, \TJ{}{'}, have been estimated to be $\sim 2000$~K from magnetic susceptibility measurements \cite{Motoyama:PRL76:1996:3212} 
and to be as ferromagnetic as $\sim -200$~K from the theoretical calculation \cite{Rice:EL23:1993:445}, respectively. 
Therefore, the zigzag spin-chain in \TSCO{}{}{2} can be regarded as two independent 1D AF spin-chains, 
but these are weakly coupled via \TJ{}{'} with frustration \cite{Matsuda:JMMM140:1995:1671}. 

Spin defects in \TSCO{}{}{2} appear on account of the vanishment of spins on {\TCuII} ions, 
that is caused by substitution of nonmagnetic impurities for Cu and hole/electron doping. 
A spin-chain in \TSCO{}{}{2} is disconnected due to defects of oxygens in the Cu-O-Cu bonds also. 
Therefore, 
long spin-chains maybe prepared in \TSCO{}{}{2} by the single-crystal growth using raw materials with high purity and by the control of the oxygen content. 
Accordingly, 
in order to enhance \Tk{spinon} in \TSCO{}{}{2}, we have grown the single crystals using raw materials with 99.9~\% (3N) and 99.99~\% (4N) purity, and annealed those in O$_2$ atmosphere 
and measured the thermal conductivity. 

%
%
\section{Experimental}
Single crystals of \TSCO{}{}{2} were grown by the traveling-solvent floating-zone (TSFZ) method. 
Polycrystalline feed rods for the TSFZ growth were prepared from SrCO$_3$ and CuO powder with 99.9~\% (3N) and 99.99~\% (4N) purity. 
It has been found from the iodine titration that as-grown single-crystals had oxygen defects. 
Therefore, as-grown single-crystals were annealed in O$_2$ gas flow at 870{\TdegC} for 48 h 
although it has been reported by Motoyama {\it et al.} \cite{Motoyama:PRL76:1996:3212} that spin defects are decreased by Ar-annealing.
Thermal conductivity measurements were carried out by the conventional steady-state method. 

%
%
\section{Results and Discussion}

\begin{figure}[t]
\begin{center}
		\includegraphics[width=18pc]{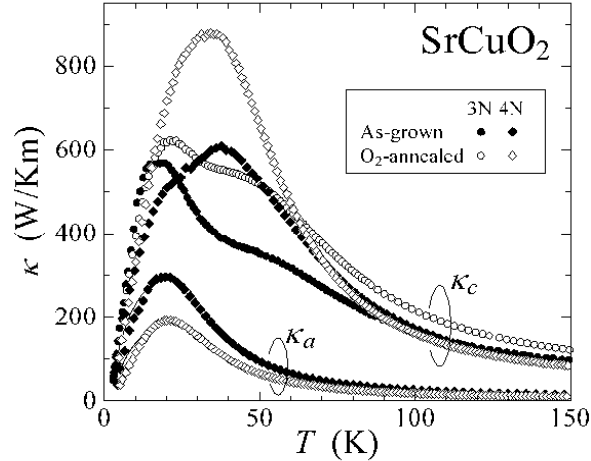}
		\caption{Temperature dependence of the thermal conductivity along the $c$-axis parallel to spin-chains, \Tkax{c}, 
and along the $a$-axis perpendicular to spin-chains, \Tkax{a}, 
for as-grown and O$_2$-annealed single-crystals of \TSCO{}{}{2} grown from raw materials with 3N and 4N purity. Solid lines are fitting results of \Tk{phonon} in \Tkax{a}.}
		\label{fig:kappa}
		\end{center}
\end{figure}

\begin{figure}[t]
		\begin{center}
		\includegraphics[width=18pc]{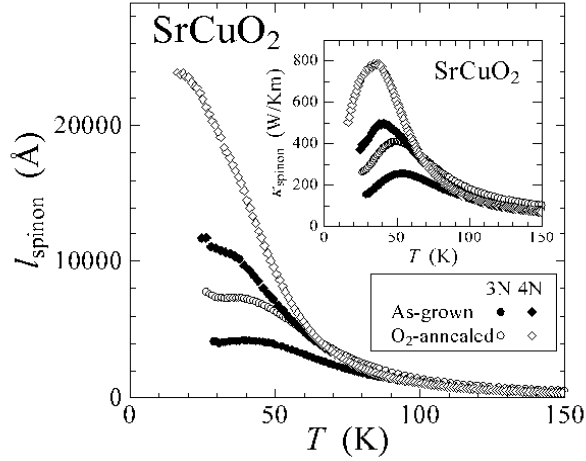}
		\caption{Temperature dependence of the mean free path of spinons, \Tl{spinon}, for as-grown and O$_2$-annealed single-crystals of \TSCO{}{}{2} grown from raw materials with 3N and 4N purity. The inset shows the temperature dependence of the thermal conductivity due to spinons, \Tk{spinon}.
		\vspace{1pc}
		}
		\label{fig:kappasp}
		\end{center}
\end{figure}

Figure 1 shows the temperature dependence of the thermal conductivity along the $c$-axis parallel to spin-chains, \Tkax{c}, 
and along the $a$-axis perpendicular to spin-chains, \Tkax{a}, 
for the as-grown and O$_2$-annealed single-crystals of \TSCO{
}{}{2} grown from raw materials with 3N and 4N purity. 
In the as-grown 3N sample, the behavior of \Tkax{c}, which exhibits a peak around $20$~K and a shoulder around $50$~K, is similar to that reported by Sologubenko {\it et al.} \cite{Sologubenko:PRB64:2001:054412}.
The peak is due to the peak of the thermal conductivity due to phonons, \Tk{phonon}, and the shoulder is due to the peak of \Tk{spinon}. 
In the as-grown 4N sample, \Tkax{c} exhibits a small shoulder around $20$~K and a peak around $40$~K. 
 Considering the peak temperature and the shoulder temperature, it is guessed that the shoulder and peak are due to peaks of \Tk{phonon} and \Tk{spinon}, respectively. 
Actually, \Tkax{a} in the as-grown 4N sample exhibits only one peak due to \Tk{phonon} around $20$~K. 
Accordingly, it is concluded that the contribution of \Tk{spinon} to \Tkax{c} is enhanced by the growth using raw materials with high purity. 

It is found that \Tkax{c} in the 3N and 4N samples is enhanced by the O$_2$-annealing also. 
In particular, the shoulder of the 3N sample and the peak of the 4N sample around $50$~K, which are due to the peak of \Tk{spinon}, are markedly enhanced by the O$_2$-annealing. 
On the other hand, \Tkax{a} of the 4N samples is suppressed by the O$_2$-annealing. 
These results mean that the number of spin defects decreases by the O$_2$-annealing which is expected to connect {\TCuII} ions separated by oxygen defects in the 180{\Tdeg} Cu-O-Cu bond and also to remove excess electrons doped by oxygen defects. 
The suppression of \Tkax{a} by the O$_2$-annealing may be caused by possible excess interstitial oxygens scattering phonons. 

Here, we estimate the contribution of \Tk{spinon} to \Tkax{c} in order to see the enhancement of \Tk{spinon} clearly. 
At first, the estimate of \Tk{phonon} is necessary. 
The temperature dependence of \Tk{phonon} is usually given by the following equation based on the Debye model \cite{TCS:1976}. 
\begin{equation}
	\Tk{phonon} = \frac{ \TkB }{2 \pi^2 \Tv{phonon}}
		\biggl( \frac{ \TkB T }{ \hbar } \biggr) ^3
		\int _0 ^{ \TDebye / T }
		\frac{x^4 e^x}{(e^x - 1) ^2}
		\Tt{phonon}
		\diff x
		, 
\label{Debye:Eq}
\end{equation}
where 
$x = \hbar \omega / \TkB T $, 
$\omega$ the phonon angular frequency, 
$\hbar$ the Planck constant, 
{\TkB} the Boltzmann constant 
and {\TDebye} the Debye temperature.
The phonon velocity, \Tv{phonon}, is calculated as 
$\Tv{phonon} = \TDebye (\TkB / \hbar) (6 \pi ^2 n) ^{-1/3}$, 
where $n$ is the number density of atoms.
The phonon scattering rate, $\tau_{\mathrm{phonon}}^{-1}$, 
is given by the sum of scattering rates due to various scattering 
processes as follows, 
\begin{equation}
	\Tt{phonon} ^{-1}
	=
	\frac{ \Tv{phonon} }{ L _{\mathrm{b}} }
	+
	D \omega
	+
	A \omega ^4
	+
	B \omega ^2 T \exp ( - \frac{ \TDebye }{ bT })
	,
\label{Tau:Eq}
\end{equation}
where $L _{\mathrm{b}}$, $D$, $A$, $B$ and $b$ are constants.
These terms represent the phonon scattering by boundaries, 
by dislocations, by point defects 
and the phonon-phonon scattering in the umklapp process in turn.
Here, {\TDebye} is put at $405.8$~K from our specific heat measurements. 
The temperature dependence of \Tkax{a} of 4N as-grown and O$_2$-annealed samples due to only \Tk{phonon} is well fitting with Eqs. (1) and (2) as solid lines in Fig. 1. 
In this estimation, values of $B$ and $b$ for 4N as-grown and O$_2$-annealing samples are determined to be the same values, respectively. 
For the fit of \Tk{phonon} to \Tkax{c}, 
values of $B$ and $b$ are put at the same values as those used for the fit to \Tkax{a}, respectively, 
because the phonon-phonon scattering in the umklapp process seems neither to be affected by the crystal direction, 
by the small change to the oxygen content, 
nor by the purity of raw materials. 
Adjusting three parameters of \Tsub{L}{b}, $D$ and $A$, \Tk{phonon} in \Tkax{c} is estimated. 
Then, \Tk{spinon} is estimated by subtracting the fitting curve of \Tk{phonon} from the data of \Tkax{c}, as shown in the inset of Fig. 2. 
The uncertain data of estimated \Tk{spinon} at low temperatures are neglected. 
In this estimation, these three parameters are determined so that \Tk{spinon} at low temperatures is proportional to temperature as theoretically expected. 
Values of the parameters used for the best fit to \Tkax{c} are listed in Table 1. 
\begin{table}[bt]
	\caption{
		Parameters used for the fit of the temperature dependence of the thermal conductivity along the $c$-axis, \Tkax{c}, with Eqs. (1) and (2).
	}
	\label{tab:fit}
	\begin{center}
	\begin{tabular}{lccccc}
		\br
		&$L_{\mathrm{b}}(10^{-3}~{\Tm})$&$D(10^{-6})$&$A(10^{-43}~{\Tsec}^3)$&$B(10^{-18}~{\Tsec}\TK^{-1})$&$b$\\
		\mr
		3N as-grown      &3.25&0.40&2.95&7.1&2.7\\
		3N O$_2$-annealed&1.55&0.80&2.25&7.1&2.7\\
		4N as-grown      &2.11&2.40&8.00&7.1&2.7\\
		4N O$_2$-annealed&2.24&6.50&8.50&7.1&2.7\\
		\br
	\end{tabular}
	\end{center}
	\vspace{-2pc}
\end{table}

Next, we estimate \Tl{spinon}, using the following equation based on the Heisenberg model and the des Cloizeaux-Pearson mode \cite{Cloizeaux:PR128:1962:2131}, 
\begin{equation}
	\Tl{spinon} = \frac{3 \hbar}{\pi \Tsub{N}{s} c \TkB ^2 T} \Tk{spinon}, 
	\label{equ:lsp}
\end{equation}
where \Tsub{N}{s} is the number of spins and $c$ is the lattice constant of the $c$-axis which is the same as the distance between the nearest neighboring spins in the spin-chain. 
It is found that \Tl{spinon} increases with decreasing temperature, as shown in Fig. 2. 
Value of \Tl{spinon} tend to be saturated at low temperatures. 
In this analysis, 
it is hard to estimate \Tk{phonon}, \Tk{spinon} and therefore \Tl{spinon} at low temperatures exactly. 
However, it is found that \Tl{spinon} is markedly extended by both the increase of the purity of raw materials and the O$_2$-annealing. 
This indicates that \Tl{spinon} is extended owing to the decrease of spin defects caused by impurities in raw materials and by oxygen defects. 
The maximum value of \Tl{spinon} in the O$_2$-annealed 4N sample is as large as $\sim 24000$~{\AA}. 
The length between spin defects calculated from the purity of raw materials, \Tsub{L}{purity}, is 3918 {\AA} and 39180 {\AA} in 3N and 4N samples, respectively. 
In the 3N sample, the maximum of \Tl{spinon} is larger than the value of \Tsub{L}{purity}. 
It is possible that the spinons carry heat over the spin defects, 
because the spinons may be able to pass the spin defects through the weak interaction \TJ{}{'} in the double spin-chains. 
This is different from the case in \TSCO{2}{}{3}. 
There is another possibility that the quality of the raw materials is higher than just 99.9~\%. 
In the 4N sample, on the other hand, the maximum of \Tl{spinon} is smaller than the value of \Tsub{L}{purity}. 
This indicates that \Tk{spinon} is disturbed by spin defects due to oxygen defects. 
Therefore, it is expected that \Tk{spinon} is enhanced by more suitable O$_2$-annealing.

%
%
\section{Summary}
We have measured the thermal conductivity of single crystals of \TSCO{}{}{2} grown from raw materials with 3N and 4N purity. 
The magnitude of \Tk{spinon} has been found to become large by both the increase of the purity of raw materials and the O$_2$-annealing. 
Accordingly, we have succeeded in the enhancement of \Tk{spinon}. 
This result indicates that \Tl{spinon} is able to be extended by the removal of spin defects in $S=1/2$ 1D AF systems where the thermal conduction due to spinons is ballistic, leading to the enhancement of \Tk{spinon}. 

%
%
\ack
This work was supported by a Grant-in-Aid for Scientific Research from the Ministry of Education, Culture, Sports, Science and Technology, Japan. 

%
%
\section*{References}

\end{document}